\documentclass[prd,a4paper]{revtex4}

\usepackage{amsmath}
\usepackage{epsfig}

\newcommand{\A}{{\mathcal A}}

\newcommand{\Td}{{\mathcal T}_{\mathrm{d}}}
\newcommand{\Ad}{{\mathcal A}_{\mathrm{d}}}
\newcommand{\as}{\alpha_s}
\newcommand{\asb}{\bar\alpha_s}

\begin{document}
\title{Geometric scaling in exclusive processes}
\author{St\'ephane Munier}
\affiliation{Institute for Theoretical Physics,
University of Heidelberg,
Philosophenweg 19, 69120 Heidelberg, Germany.}
\email{munier@tphys.uni-heidelberg.de}
\author{Samuel Wallon}
\affiliation{Laboratoire de Physique Th\'eorique, Universit\'e Paris
  XI, centre d'Orsay, b\^atiment 211, 91405 Orsay cedex,
France.}
\email{Samuel.Wallon@th.u-psud.fr}

\begin{abstract}
We show that according to the present understanding of the energy
evolution of the observables measured in deep-inelastic scattering,
the photon-proton scattering amplitude
has to exhibit geometric scaling at each impact
parameter. We suggest a way to test it experimentally at HERA.
A qualitative analysis based on published data
is presented and discussed.
\end{abstract}

\maketitle

\section{Introduction}

The striking phenomenon of geometric scaling discovered in the HERA
data \cite{Stasto:2000er} has triggered a wide interest.
The inclusive deep-inelastic scattering (DIS) cross section $\sigma(x,Q)$
measured as a function of the photon virtuality $Q$ and the Bjorken
variable $x$, was shown to depend on the single combined variable 
$Q^2\cdot (x/x_0)^{\lambda_0}$ where
$\lambda_0\!\sim\!0.3$ and $x_0\!\sim\!3\cdot 10^{-4}$. 
Empirically, this scaling turns out to be
valid for $x\!\le\!10^{-2}$ and for all available values of $Q^2$ (in
practice, for $Q^2\!=\!0$ to $400\ \mbox{GeV}^2$).
Geometric scaling has also been found more recently
in deep-inelastic scattering off
nuclei~\cite{Freund:2002ux}.

A similar scaling behaviour was discussed a long time ago in the framework of
elastic proton-proton scattering \cite{Buras:1974km,Amaldi:1980kd}. At that time, 
an impact parameter analysis revealed
that the rise of the cross section as a power of the center-of-mass energy was
due to an effective growth of the size of the interaction region,
which was almost black in its center.
Geometric scaling in $\gamma^*-p$ scattering 
was discovered in the context of the discussion of saturation effects,
and is often considered as an evidence for the fact that the
proton looks already quite black, or ``saturated'', at HERA.

The scaling variable turns out to be the dimensionless 
ratio of the photon virtuality $Q$ and the
so-called saturation scale, which has the
following dependence upon the Bjorken variable:
\begin{equation}
Q_s(x)\sim 1\;\mbox{GeV}\cdot\left(\frac{x}{x_0}\right)^{-\lambda_0/2}\ .
\label{eq:satscal1}
\end{equation}
The latter is believed to characterize the 
mean transverse momentum of partons with longitudinal momentum
fraction $x$.
The growth of this typical momentum scale with energy $W\!=\!Q/\sqrt{x}$
is understood within several different formulations of dense parton
evolution \cite{Balitsky:1996ub,Kovchegov:1999yj,
McLerran:1994ni,McLerran:1994ka,McLerran:1994vd,Iancu:2001md,
Iancu:2000hn,Ferreiro:2001qy}
which technically lead to non-linear integro-differential equations.
Geometric scaling was 
seen in numerical simulations of these models
\cite{Lublinsky:2001bc,Golec-Biernat:2001if}.
It was shown to be consistent with QCD evolution in the range in which
it is verified in the data \cite{Kwiecinski:2002ep}.
In Ref.\cite{Iancu:2002tr,Mueller:2002zm,Triantafyllopoulos:2002nz},
it was derived from perturbative QCD for $Q$ above the saturation scale
as a feature of the Balitsky-Fadin-Kuraev-Lipatov (BFKL) equation 
\cite{Kuraev:1977fs,Balitsky:1978ic} with appropriate boundary conditions.
In Ref.\cite{Munier:2002gf}, 
the symmetry of the quantity $Q/Q_s\cdot\sigma(Q/Q_s)$ under the interchange 
of $Q$ and $Q_s$ already noticed in Ref.\cite{Stasto:2000er}
was interpreted as a possible evidence for a
proton looking like a collection of independent dipoles of
typical size $1/Q_s$.
The generalized vector dominance model combined with the color dipole
model as formulated in \cite{Kuroda:2002ry}
also leads to geometric scaling.
More recently, geometric scaling for all $Q^2$ was deduced from the critical behaviour
of the correlation function of Wilson lines
which appears at small $x$ in a near-light-cone 
Hamiltonian formalism \cite{Pirner:2002us}. This approach looks
fundamentally different from the other ones, in the sense that it
does not rely on the parton model. 

Most of the latter discussions were only concerned with inclusive
quantities. The proton was assumed an homogeneous disk,
and its size large with respect to the size of the probe, so that
boundary effects could be neglected.
The reason for these assumptions is the technical complication of 
the saturation equations when full dependence upon transverse coordinates is taken
into account.

However, one knows that the saturation scale itself depends on
the impact parameter. 
Indeed, it is related to the density of partons inside the proton,
which obviously decreases smoothly when
one moves from the center to the periphery.
The saturation scale in the ideal case described before was only a kind of effective
one. Actually, it is theoretically not clear what this scale is in a
real proton, because all momenta from $\Lambda_{\mathrm{QCD}}$
to $Q$ are {\it a priori} expected to contribute to cross sections.

The importance of impact parameter studies for the phenomenology of saturation
effects at HERA was explained in Ref.\cite{Munier:2001nr}. A
theoretical discussion of the impact parameter dependence of the
saturation scale $Q_s(x,b)$ was provided in~\cite{Ferreiro:2002kv}.
In the latter paper, some arguments were put forward in
favor of a ``local'' geometric scaling, which would manifest itself in
the fact that cross sections for exclusive processes only depend on
the ratio of scales $Q/Q_s(x,b)$.
Let us also stress that geometric scaling for exclusive processes has
recently been predicted qualitatively in the context of the discussion of generalized
parton distributions \cite{Freund:2002ff}.

The aim of this paper is to show that ``local'' geometric scaling
is testable experimentally.
In the first section, we explain why this new scaling 
is expected.
We then propose a method to investigate it at HERA. Finally, we discuss the
qualitative results
obtained so far by applying our method to the data.

\section{Saturation and geometric scaling}

In this section, we show how local scaling appears within different
theoretical approaches. We consider the idealized process of
scattering of a dipole of size $1/Q$ off a dipole of size $1/\Lambda$.
$\Lambda$ and $Q$ are fixed momentum scales, which are taken in the perturbative
region in the following discussion, and the ordering $Q>\Lambda$ is
assumed. The small dipole is called ``projectile'' and the large one
``target''. In the end, $\Lambda$ will
be formally extrapolated to $\Lambda_{{\mathrm{QCD}}}$.
We denote by $y\!\equiv\!\log(1/x)$ the maximum rapidity available at
c.o.m. energy $W\!=\!Q/\sqrt{x}$.

\subsection{Global and local scaling above the saturation scale from linear evolution}

We explore here the perturbative picture of high energy scattering.
We consider dipole-dipole scattering at fixed impact parameter $b$ as
given by the BFKL equation.
We assume that $b$ is larger than the sizes of the initial-state
dipoles: $b\!\gg\!1/Q, 1/\Lambda$.
The Mellin representation of the solution of the BFKL equation
then reads \cite{Navelet:1998tx,Munier:1998nj}
\begin{equation}
\Ad(y,Q,b)=\frac12\frac{1}{Q^2 \Lambda^2 b^4}
\int\frac{d\gamma}{2i\pi}(1\!-\!2\gamma)d(\gamma)
\left({16 b^2}{Q\Lambda}\right)^{2\gamma}
e^{\asb y \chi(\gamma)}\ ,
\label{eq:ampb}
\end{equation}
where the integration goes over a line in the complex plane 
parallel to the imaginary
axis and intersecting the real axis between 0 and 1.
In our conventions, the $S$ and $T$ matrices are
related by $S=1\!-\!T$, so that generally speaking 
the amplitudes are essentially real at
high energy. This is true in particular for the BFKL solution~(\ref{eq:ampb}).
The following standard notations have been used:
$\asb=\as N_c/\pi$ and
$\chi(\gamma)\!=\!2\psi(1)\!-\!\psi(\gamma)\!-\!\psi(1\!-\!\gamma)$.
The function  $d(\gamma)\!=\!\as^2/(16\gamma^2(1\!-\!\gamma)^2)$
is the Mellin transform of the elementary dipole-dipole elastic
amplitude with respect to the ratio of their sizes, with appropriate normalization.

The forward amplitude $\Td$ is obtained by integrating~(\ref{eq:ampb}) over the impact
parameter $b$. As formula~(\ref{eq:ampb}) holds only in a large $b$
approximation, we must set the lower bound of the integral to a number of the order of
the size of the largest interacting dipole. 
We choose $1/(4\Lambda)$ and check the consistency a posteriori.
The forward amplitude then reads
\begin{equation}
\Td(y,Q)=2\pi\int_{1/(4\Lambda)}^\infty db\,b\,\Ad(y,Q,b)
 =\frac{8\pi}{Q^2}\int\frac{d\gamma}{2i\pi}d(\gamma)
\left(\frac{Q}{\Lambda}\right)^{2\gamma} e^{\asb y \chi(\gamma)}\ .
\label{eq:forward}
\end{equation}
One checks the normalizations by computing the total cross section,
using the optical theorem which reads $\sigma\!=\!2\Td$ in our conventions.
Setting $y=0$, one recovers the well-known expression of the
elementary (lowest order) dipole-dipole cross section~\cite{Mueller:1994jq}.

Let us recall how one obtains geometric scaling from this equation.
One is interested in the large $y$ region, where a saddle point
gives the dominant contribution to the integral
in~(\ref{eq:forward}). 
Expanding the function $\chi(\gamma)$
to second order around $\gamma\!=\!1/2$, the amplitude~(\ref{eq:forward}) reads
\begin{equation}
\Td(y,Q)=\frac{4\pi\as^2}{\sqrt{\pi a^2 y}}\frac{1^{\phantom{2}}}{\Lambda^2}
\exp\left\{\omega y -\frac12\log(Q^2/\Lambda^2)-\frac{1}{4a^2
    y}\log^2(Q^2/\Lambda^2)\right\}\ ,
\label{eq:saddleinter}
\end{equation}
with the definitions $\omega\!\equiv\!4\asb\log 2$ (the BFKL intercept) 
and $a^2\!\equiv\!14\zeta(3)\asb$.
At fixed $y$, the amplitude becomes large (of the order of the area of
the target) when the argument of the exponential is positive, which occurs
below the scale $Q\!=\!Q_s(y)$ satisfying the relation
\begin{equation}
Q_s^2(y)=\Lambda^2 e^{\lambda y},\ \mbox{with}\ 
\lambda\!\equiv\!a^2\left(\sqrt{1+4\omega/a^2}-1\right)\ .
\label{eq:satscal}
\end{equation}
Note that this formula can be compared with the form of the
saturation scale previously quoted~(\ref{eq:satscal1}) provided one
makes the identifications $\lambda\rightarrow\lambda_0$ and
$\Lambda\rightarrow 1\ \mbox{GeV}\cdot x_0^{\lambda_0/2}$.
The equation $Q^2\!=\!Q_s^2(y)$ defines the so-called critical line
below which, per definition, 
the amplitude~(\ref{eq:forward}) is large and non-linear corrections
prove to be necessary to ensure probability conservation.
The interesting feature of Eq.(\ref{eq:saddleinter}) is that it can be rewritten
as a function of the ratio of the variables $Q$ and $Q_s$, up to some
approximation.
Putting~(\ref{eq:saddleinter}) and~(\ref{eq:satscal}) together, one
arrives at
\begin{equation}
\Td(y,Q)=\frac{4\pi\as^2}{\sqrt{\pi a^2 y}}
\frac{1}{\Lambda^2} \exp\left\{
-\frac{1}{a^2 y}\log^2\frac{Q}{Q_s(y)} 
\right\}
\left(\frac{Q}{Q_s(y)}\right)^{-\sqrt{1+4\omega/a^2}}\ .
\end{equation}
The factor which still contains the rapidity alone is slowly varying, hence it
can be approximated by its value for a fixed typical rapidity $y_0$.
Moreover,
if $Q$ is sufficiently close to $Q_s$, one can trade the exponential
for $1$.
This approximation applies when $|\log Q/Q_s|\!\ll\!a^2 y \sqrt{1+4\omega/a^2}$.
Introducing the notation 
$\gamma\!\equiv\!(\sqrt{1\!+\!4\omega/a^2})/2$, the amplitude rewrites
\begin{equation}
\Td(Q,Q_s(y))=\frac{4\pi\as^2}{\sqrt{\pi a^2 y_0}}
\frac{1}{\Lambda^2}
\left(\frac{Q_s^2(y)}{Q^2}\right)^{\gamma}\ ,
\label{eq:geof}
\end{equation}
which exhibits geometric scaling.

To make contact with HERA phenomenology, one has to set
$\Lambda\!=\!\Lambda_{\mathrm{QCD}}$
and to average the projectile size $r$ with weight given by
the squared photon wave function. This is the well-known dipole model
\cite{Nikolaev:1991ja,Mueller:1994rr} which provides an appealing and useful picture of
DIS in the leading-logarithmic approximation of QCD.
The virtual photon-proton
cross section reads
\begin{equation}
\sigma_{\gamma^* p}(y,Q)=\int d^2 r\, \psi^{\dagger}_{\gamma^*}(r,Q)
\otimes\psi_{\gamma^*}(r,Q)
\cdot2\Td(y,1/r)\ .
\label{eq:conv}
\end{equation}
$\psi_{\gamma^*}$ is the photon wave function on a $q\bar q$ colour dipole.
Besides the transverse size $r$ and virtuality of the photon $Q$, this
wave function also depends on  the longitudinal
momentum fraction $z$ of the antiquark.
The notation~``$\otimes$'' stands for an integration over $z$.
For negligible quark masses, $\psi^{\dagger}_{\gamma^*}(r,Q)
\otimes\psi_{\gamma^*}(r,Q)$ is a function of $rQ$ only
which vanishes rapidly at large $rQ$ \cite{Stasto:2000er}.
Thus the scaling property of the dipole-proton amplitude~(\ref{eq:geof}) is
transmitted to the photon-proton cross section itself which then
exhibits the same behaviour as the amplitude $\Td$ in Eq.(\ref{eq:geof}).

The formula just established 
would give a reasonable description of the data in a limited kinematical
range, provided one considers $\omega$ and $a^2$ as effective
parameters.
Indeed, the values given by the leading order BFKL equation are
known to be incompatible with the data on one hand, and to receive large
corrections from next-to-leading order terms on the other hand.
The parameters have to be adjusted
such that $\lambda\sim\lambda_0\sim 0.3$ and $\gamma\sim 1$ 
\cite{Golec-Biernat:1999qd,Munier:2002gf}.
Then formulae~(\ref{eq:satscal1}), (\ref{eq:satscal}) show that the
choice $x_0\!=\!3\cdot 10^{-4}$ implies $\Lambda_{\mathrm{QCD}}=300\
\mbox{MeV}$, which is reasonable.
To obtain more reliable expressions, neglected effects such as
diffusion into the infrared have to be taken into account. As shown in
Ref.\cite{Mueller:2002zm}, the latter have sizable effects both on the
parametric form of the
saturation scale~(\ref{eq:satscal}) and on the expression of 
the amplitude~(\ref{eq:geof}), but do not
spoil the overall picture.

The previous analysis can be taken over to the impact parameter
dependent amplitude~(\ref{eq:ampb}) in a straightforward way.
The steepest-descent method gives (see \cite{Kovner:2001bh})
\begin{equation}
\Ad(y,Q,b)=\frac{128\pi\as^2}{(\pi a^2 y)^{3/2}}
\log(16 b^2 Q\Lambda) \exp\left\{\omega y-\log(16 b^2 Q\Lambda)
-\frac{1}{a^2 y}\log^2(16 b^2 Q\Lambda)\right\}\ .
\end{equation}
This amplitude is again of order $1$ for $Q\!=\!Q_s$, the saturation
scale $Q_s$ being given by
\begin{equation}
Q_s^2(y,b)=\frac{1}{256 \Lambda^2 b^4} \exp(\lambda y)\ .
\label{eq:satscal2}
\end{equation}
Within the approximations already made in the inclusive case,
the amplitude can then be expressed as a function of $Q$ and $Q_s$ only:
\begin{equation}
\Ad(Q,Q_s(y,b))=\frac{64\as^2}{\sqrt{\pi a^2 y_0}}
(2\gamma\!-\!1)\cdot\left(\frac{Q_s^2(y,b)}{Q^2}\right)^{\gamma}\ ,
\label{eq:scalingfinal}
\end{equation}
where the saturation scale now depends on the impact parameter $b$. 
Thus we see that 
the amplitude for each impact parameter is a
function of the ratio of the inverse size of the projectile and of
the {\it local} saturation scale.

However, the kinematical range in which formula~(\ref{eq:scalingfinal}) may apply is
quite restricted. 
Let us set $\Lambda=\Lambda_{\mathrm{QCD}}$.
We have required large impact
parameters (in practice, $b>0.2\ \mbox{fm}$)
and $Q_s$ close to $Q$, which has to sit in the
perturbative regime ($Q\gg\Lambda_{\mathrm{QCD}}$).
From formula~(\ref{eq:satscal2}), one sees that the latter
conditions require a large rapidity evolution, during which the
saturation regime might already be reached for smaller impact
parameters (i.e. $Q<Q_s(b^{\prime})$ for some $b^{\prime}<b$). 
It is not possible to quantify this effect without knowing the
dynamics for all impact parameters.

We shall now refer to another, maybe safer approach, which will
however lead to comparable conclusions on ``local'' geometric scaling.

\subsection{The impact parameter profile as an initial condition}

In Ref.\cite{Ferreiro:2002kv}, the argument that the quantum evolution was
quasi-local in impact parameter was developed on the basis of an
analysis of the non-linear evolution equations.
As a result, the dipole-proton scattering amplitude at fixed impact
parameter can be written in
a factorized form, because the latter feature means that
the impact parameter dependence remains unchanged
throughout the rapidity evolution. The non-perturbative profile of the
proton is introduced as an initial condition.
The validity of such a factorization
has been confirmed very recently in the context of $\gamma^*-\gamma^*$
reactions \cite{Bondarenko:2003ym}.
The amplitude then reads
\begin{equation}
\Ad(y,Q,b)=S(b)\cdot \Td(y,Q)\ ,
\label{eq:rapevol}
\end{equation}
where $\Td(y,Q)$ is the forward amplitude~(\ref{eq:forward}).
Above the saturation scale, $\Td(y,Q)$ exhibits geometric scaling
as seen on Eq.(\ref{eq:geof}),
thus the amplitude $\Ad$ also scales.
The profile function
\begin{equation}
S(b)=\frac{2}{\pi} m_{\pi}^2\cdot e^{-2m_{\pi} b}
\end{equation}
was postulated (we chose the normalization such that $\int d^2b\,S(b)\!=\!1$).
It involves the long-distance scale $1/m_{\pi}$, which is put by hand
and interpreted as an initial condition.

This leads once again to a scaling of the
form~(\ref{eq:scalingfinal}), with a local saturation scale
given by
\begin{equation}
Q_s^2(y,b)=e^{-2m_{\pi}b/\lambda}\cdot Q_s^2(y)
=e^{-2m_{\pi}b/\lambda}\cdot \Lambda_{\mathrm{QCD}}^2 e^{\lambda y} \ .
\label{eq:satval2}
\end{equation}
We see that this result is qualitatively close to Eq.(\ref{eq:satscal2}), except for
the $b$-dependence which is exponential rather than powerlike.
This stems from the fact that non-perturbative physics with short
range pion fields were introduced here, whereas previously only
long-range Coulomb-like fields were considered.

At this point, it may be useful to 
evaluate the saturation scale at HERA from these formulae.
For values of the rapidity $y$ 
of the order of those measured at HERA in the small-$x$ region
($x\!\sim\! x_0\!\sim\!3\cdot 10^{-4}$), 
one sees from formula~(\ref{eq:satval2}) that 
$Q_s(b\!=\!0)\!\sim\! 1\ \mbox{GeV}$.
$Q_s$ reaches the value of $\Lambda_{\mathrm{QCD}}$ at the impact
parameter $b\!\sim\!0.6\ \mbox{fm}$.
This means that for a probe of size $1/Q\!<\!1\ \mbox{GeV}^{-1}$, the
proton looks grey everywhere and extends up to a radius of about $0.6\ \mbox{fm}$.

\section{A way to check scaling at HERA}

As geometric scaling at fixed impact parameter seems to be a definite
prediction of saturation models, it is worth testing it against the data.
A former analysis \cite{Munier:2001nr} has already shown the relevance of impact parameter
dependent analysis in the discussion of saturation. There, the
$S$-matrix element for dipole-proton scattering at fixed impact
parameter was extracted from diffractive electro-production of
vector mesons at HERA. This quantity can be interpreted as a transparency
coefficient, and thus quantifies the ``blackness'' of the proton as seen by the
projectile.
Such kind of analysis is always model-dependent, since to get
the correct normalization of the $S$-matrix element, one has to rely
on an {\it ad hoc}  model for the final-state vector meson.
Although it was shown in Ref.\cite{Munier:2001nr} that requiring that
the initial state be a {\it longitudinal} photon limits the model-dependence, 
an estimated uncertainty of 20\% was still recognized.
However, dependence of the cross section on a scaling variable
can be tested in a model-independent way.

The scattering amplitude at fixed impact parameter can be extracted from
differential cross sections for high energy quasi-elastic processes
by a Fourier transform with respect to the momentum transfer. 
By quasi-elastic we mean that the initial and final states
have the same number of particles. In practice, the process we are
thinking of will be diffractive electroproduction of vector mesons.

Let us consider such a quasi-elastic process. Its amplitude $\A$ is
related to the differential cross section through the formula
\begin{equation}
\frac{d\sigma}{dt}=\frac{1}{16\pi}|\A(t)|^2\ ,
\label{eq:differential}
\end{equation}
where $t$ is
the usual Mandelstam variable and all other dependences
are implied. At high energies, it is well-known that the scattering
amplitude $\A$ is essentially real.
Formula~(\ref{eq:differential}) shows that $\A(t)$ is proportional to
the square root of the differential cross
section $d\sigma/dt$.
A further Fourier transform
with respect to the two-dimensional momentum transfer then
gives the scattering amplitude for a fixed impact parameter
\begin{equation}
\A(b)=\sqrt{\pi}\int_{0}^{+\infty}d(-t)\,J_0(b\sqrt{|t|})\sqrt{\frac{d\sigma}{dt}}\ ,
\label{eq:transfo}
\end{equation}
where $J_0$ is a Bessel function.
For formula~(\ref{eq:transfo}) to be valid, it is important that
the quasi-elastic final state be well identified as a {\it single} particle
or resonance. A more complex final state like 
$\gamma^*p\rightarrow\pi^+\pi^- p$ (which is what is actually observed
in the experiments) can proceed through several different
channels (with intermediate states $\rho$, $\omega$, $f_0$...). Its study
would require the careful treatment of phases and
interference effects, see for example Refs.\cite{Hagler:2002nh,Hagler:2002nf}.
We shall assume that we have a single-channel process, and 
we use formula~(\ref{eq:transfo}) to extract the $b$-dependent
amplitude from the available data.

We shall now be more specific, specializing 
to the case of electroproduction of vector mesons
$\gamma^*p\rightarrow Vp$, where $V$ will in practice be a $\rho$ meson.
The picture is the following.
At high energy, the photon splits into a dipole of fixed size $r$ which scatters off the
proton before recombining into a meson. 
Due to its high relative velocity, the dipole does not change size
during the scattering.
Accordingly, the 
scattering amplitude $\A$ is the convolution of
the photon wave function $\psi_{\gamma^*}$, the meson wave-function
$\psi_V$ and the dipole amplitude $\Ad$ in the same way as in Eq.(\ref{eq:conv})
\begin{equation}
\A(y,Q,b)=\int d^2r\,
\psi^\dagger_{\gamma^*}(r,Q)\otimes\psi_V(r)\cdot \Ad(y,1/r,b)\ ,
\label{eq:ampinter}
\end{equation}
the forward amplitude $2\Td$ appearing in~(\ref{eq:conv}) being
replaced by the impact parameter dependent one $\Ad$.
Due to the fact that the product of the wave functions 
$\psi^\dagger_{\gamma^*}\otimes\psi_V$ is peaked
around a typical dipole size depending only on the external transverse
momentum scales, the scaling of the dipole
amplitude $\Ad$ (see Eqs.(\ref{eq:scalingfinal})) is transmitted to the photon-proton amplitude $\A$.
As it is well-known from both empirical tests
\cite{Nemchik:1994fp,Nemchik:1997cw} and experimental \cite{Marage:2001sy}
studies, the relevant distance scale at the photon-meson vertex is a
combination of the photon virtuality $Q$ and of the meson mass $M$, namely
$1/\sqrt{Q^2\!+\!M^2}$ instead of $1/Q$ as in the case of inclusive scattering.
We then obtain from Eqs.(\ref{eq:satscal2}),
(\ref{eq:satval2}) and~(\ref{eq:scalingfinal}) 
that the amplitude $\A$ in Eq.(\ref{eq:ampinter}) should be a function of the
scaling variable $\tau\!=\!(Q^2\!+\!M^2)(x/x_0)^\lambda$ only, once
the impact parameter $b$ is fixed.

\begin{figure}
\epsfig{file=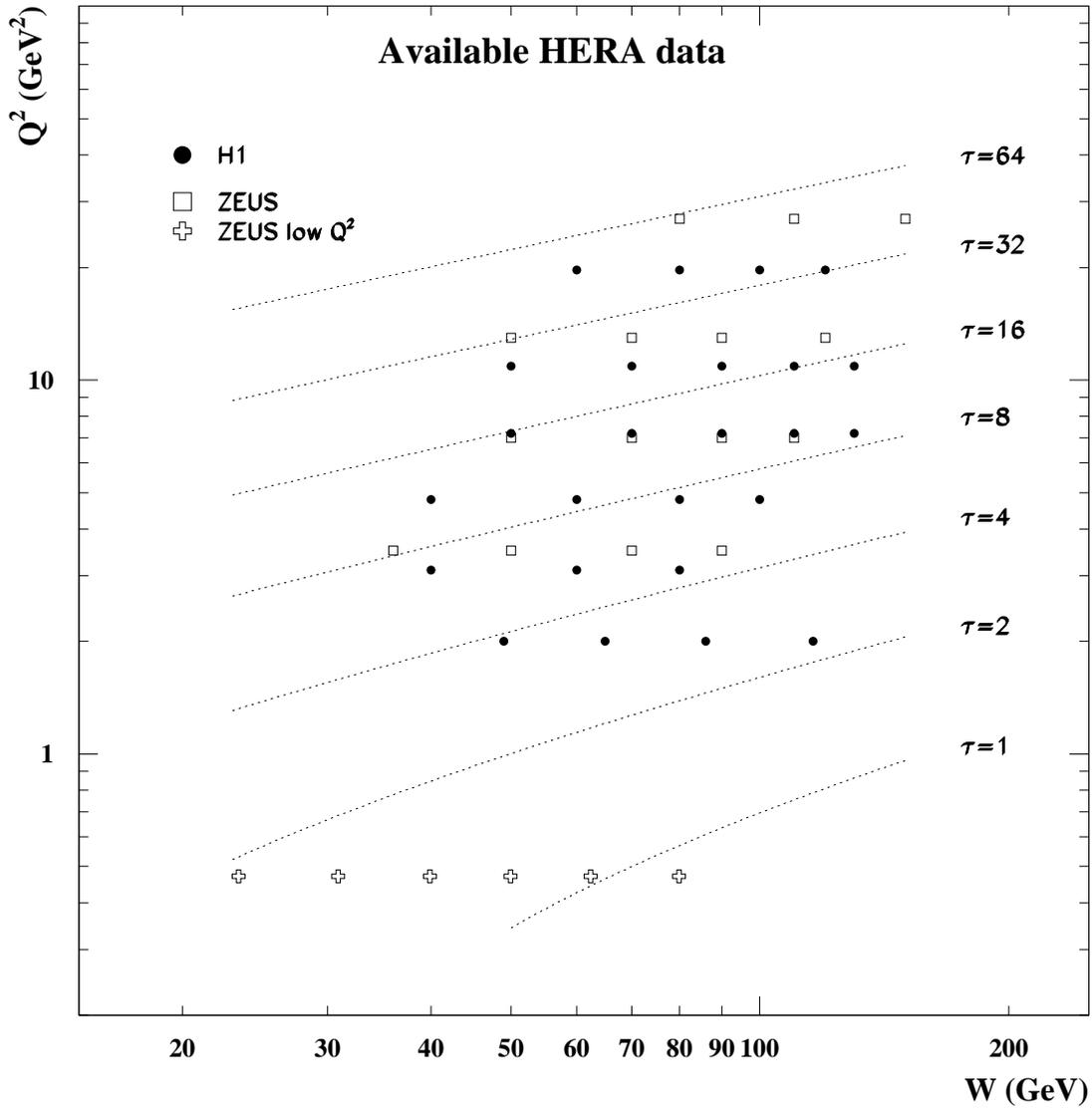,width=16cm}%
 \caption{\label{fig:kin}Kinematics of the data
on the total cross section for diffractive 
electroproduction of $\rho^0$ mesons at HERA \cite{Adloff:1999kg,Breitweg:1998nh}.
The points represent the location of the data in the kinematical plane $(Q^2,W)$.
Curves of equal scaling ratio $\tau=(Q^2+M^2)(x/x_0)^{\lambda}$ are shown.}
\end{figure}

To check this statement, we apply the
transformation~(\ref{eq:transfo}) to the HERA published 
data on electroproduction of $\rho^0$ vector mesons.
In the presently published experimental analysis \cite{Breitweg:1998nh,Adloff:1999kg},
the total cross section as well as the logarithmic
$t$-slope $B$ are quoted.
The data can then be represented by the parametrization
\begin{equation}
\frac{d\sigma}{dt}(Q,W)=B(Q,W)\cdot \sigma(Q,W)\cdot e^{-B(Q,W)|t|}\ ,
\end{equation}
up to $|t|$ of the order of $|t_{\mathrm{max}}|=1\ \mbox{GeV}^2$, which corresponds
to the range in which data have been collected.
The scattering amplitude at each impact parameter
reads
\begin{equation}
\begin{split}
\A(y,Q,b)&=\sqrt{\pi}\left(\int_0^{|t_{\mathrm{max}}|}
d(-t)\, J_0(b\sqrt{|t|})
\sqrt{B\sigma}\cdot e^{-B|t|/2}
+\int^{+\infty}_{|{t_{\mathrm{max}}}|}
d(-t)\, J_0(b\sqrt{|t|})\sqrt{\frac{d\sigma}{dt}}\right)\\
&\simeq
\sqrt{\pi}\left(\int_0^{+\infty}
d(-t)\,J_0(b\sqrt{|t|}) \sqrt{B\sigma}\cdot e^{-B|t|/2}\right)
=
2\sqrt{\frac{\pi\sigma(Q,W)}{B(Q,W)}} e^{-b^2/2B(Q,W)}\ .
\end{split}
\label{eq:expo}
\end{equation}
We take the total cross section from
\cite{Breitweg:1998nh,Adloff:1999kg}.
The kinematical range of these measurements is shown on Fig.\ref{fig:kin}.
The experimental points range from $Q^2=0.47$ to $27\ \mbox{GeV}^2$
and $W=23.4$ to $150\ \mbox{GeV}$.
Due to the lack of data for the slope $B$ for all values of $Q^2$
and $W$, we shall assume
\begin{equation}
B(Q,W)=0.6\cdot\left(\frac{14}{(Q^2+M^2)^{0.26}}+1\right)
+4\alpha^\prime\log(W/75)\ ,
\label{eq:B}
\end{equation}
all quantities 
appearing in the previous formula
being expressed in powers of 1~GeV.
The parametrization~(\ref{eq:B}) at $W=75\ \mbox{GeV}$ 
is taken from~\cite{Caldwell:2001ky}.
A logarithmic energy dependence has been added according to the
Donnachie-Landshoff parametrization for the Pomeron~\cite{Donnachie:1984xq}. We set the
Regge slope $\alpha^{\prime}$
to the value $0.25$ which is measured in hadron-hadron cross sections.
The relevance of this choice and its implications will be discussed in
the next section.

\begin{figure}
\epsfig{file=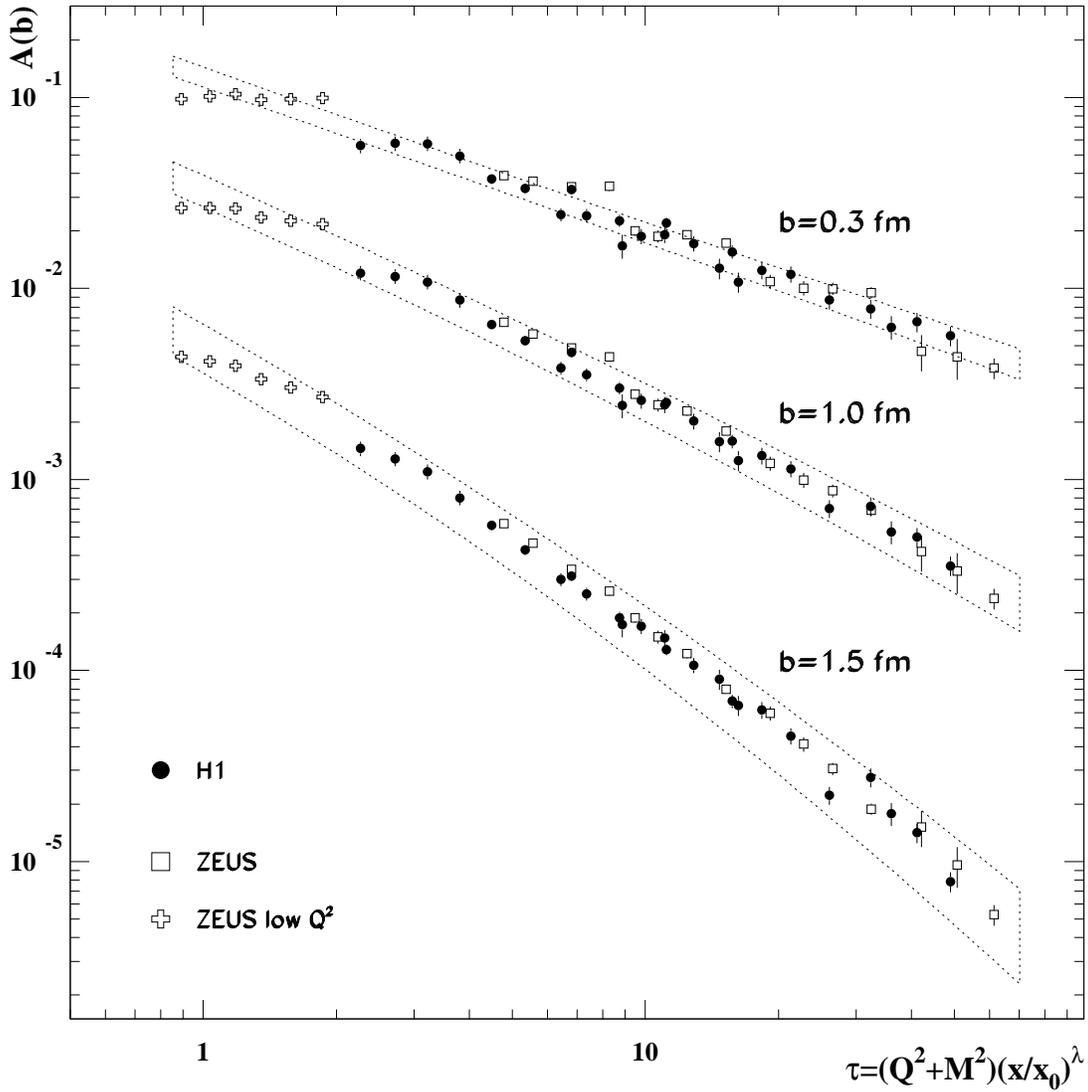,width=16cm}%
 \caption{\label{fig:local}Amplitude for 3 different impact parameters
as a function of the scaling variable.
The dotted contours indicate a rough estimate of the total uncertainty
due to experimental error on both the total cross section $\sigma$ and the
slope $B$ as explained in the text, see
formula~(\ref{eq:uncertainty}). 
The data points are derived from H1 \cite{Adloff:1999kg} and ZEUS
\cite{Breitweg:1998nh} analysis of diffractive production of $\rho^0$
mesons. The error bars shown take into account the uncertainty on the
measurement of the total cross section only.}
 \end{figure}

The results of our analysis 
are displayed on Fig.\ref{fig:local} for three different impact
parameters, $b\!=\!0.3,1.0\ \mbox{and}\ 1.5\ \mbox{fm}$.
Needless to say, our plot is only illustrative of the method and, being
theoreticians, we cannot provide a quantitative analysis.
The error bars shown take into account only the error on the total
cross section ($\Delta\A/\A\!=\!(\Delta\sigma/\sigma)/2$) and
the contour gives a rough idea of the total uncertainty, evaluated as if they
were uncorrelated and small by the formula
\begin{equation}
\frac{\Delta\A}{\A}=\frac12\frac{\Delta\sigma}{\sigma}
+\frac12\frac{\Delta B}{B}\left(1+\frac{b^2}{B}\right)\ .
\label{eq:uncertainty}
\end{equation}
We assumed an error on the slope $B$ of 5\% which seems realistic
regarding the quality of recently measured data 
(see the $J/\Psi$ analysis of \cite{Chekanov:2002xi}).
We see from~(\ref{eq:uncertainty}) that the uncertainty on $B$ affects
mostly the large impact parameter region. For small $b$, an additional
error would come from the fact that the data does not extend to
arbitrarily large $|t|$, hence the exponential form~(\ref{eq:expo}) is
not constrained for small $b$. This is the reason why we performed our
analysis only down to impact parameters as large as 0.3~fm (see
\cite{Munier:2001nr} for a more complete discussion).

The results on Fig.\ref{fig:local} are consistent with ``local''
geometric scaling within uncertainties, for each of the displayed
impact parameters. 
Although the kinematical range of the data is not very large,
scaling
is a non-trivial feature since data points of different $Q$ and $W$
overlap, as seen on Fig.\ref{fig:kin}.
We stress that from the technical point of view, the fact that the
scaling holds for {\it various} values of $b$ comes from a subtle
interplay between the $(Q,W)$ dependence of $\sigma$ and $B$ in formula~(\ref{eq:expo}).
Finally, let us note that the data points denoted by a cross
correspond to very low values of $Q^2$ ($0.47\ \mbox{GeV}^2$), and
are kinematically quite separated from the other data. 
As they are measured with a different method, we
do not make a point of the fact that they do not seem to lie on the
same curve as the larger $Q^2$ data.

\section{Discussion and conclusion}

In this paper,
we have demonstrated that ``local'' geometric scaling, predicted by
the saturation models for deep-inelastic scattering, is compatible
with the HERA data.
However, a consistent experimental analysis with
careful treatment of uncertainties
as well as an extension of the kinematical range is needed to draw a strong conclusion.
In particular, an extension of the
range in $W$ would enable us to compare measurements sitting at very different
$Q^2$ but at the same scaling variable $\tau$.
This would constitute a definite test of this local scaling.

Nevertheless, let us note
the surprizing fact that scaling seems to work well
for large
impact parameters, very far from the center of the proton, and even
outside the ``grey'' zone ($b=1.5\ \mbox{fm}$), where the concept of saturation scale
makes little sense.
This feature depends of course a lot on the form of the measured $B$, and
particularly on its energy dependence, which was introduced in an {\it
  ad hoc} way in
Eq.(\ref{eq:B}). Interestingly enough, if the energy dependence were less steep
for large momentum transfers $\sqrt{|t|}$,
scaling would also be slightly better verified at small impact parameters.
Such a feature is likely to happen since it seems that the harder the momentum scales, 
the weaker the energy-dependence of the $t$-slope.
This rule of thumb is most seen in large-$t$ proton-dissociative photoproduction of
$\rho$ mesons \cite{Chekanov:2002rm}, although of course it is not a quasi-elastic process
and our analysis would not apply.
Moreover, recent preliminary experimental results indicate that
$\alpha^{\prime}$ is smaller than
the Donnachie-Landshoff value, and decreases with the scale $Q^2+M^2$.
If scaling were confirmed also for large impact parameters, 
it might be an indication of some more general symmetry of high energy dynamics.

One would in principle be able to plot all data for different $b$ as a
function of a unique scaling variable $Q^2/Q_s^2(y,b)$, see formula~(\ref{eq:scalingfinal}).
However, we checked that the parametrizations for the
saturation scale proposed so far in Eqs.(\ref{eq:satscal}), (\ref{eq:satscal2})
do not lead to a superposition of all points on a single curve.
This means that the available approaches to the computation of the impact
parameter dependence of the saturation scale are not good enough yet.

The theoretical status of the impact parameter dependence of the
saturation scale is still unclear, mainly due to technical
difficulties in solving the full saturation equations. 
Better insights into these issues would help a lot the understanding of
high energy scattering within QCD. Important issues like unitarity problems require a
good understanding of the evolution in impact parameter 
\cite{Kovner:2001bh,Iancu:2002tr} (see Ref.\cite{Shoshi:2002in} for a model which
predicts such an evolution).
On the experimental side, diffractive vector meson production in DIS
is being investigated intensively at
HERA \cite{Levy:2003qz}, and more data will be available soon.
We believe that it would be worth to plot the forthcoming data in the way we
propose in this paper.

\begin{acknowledgments}
This work is funded by the 
European Commission IHP program under contract HPRN-CT-2000-00130.
We thank H.-J. Pirner for a critical reading of the manuscript and
useful suggestions. We are also grateful to A.~Freund, E.~Iancu and B.~Pire
whose remarks helped us to improve the manuscript.
\end{acknowledgments}


\end{document}